\documentclass[useAMS,usenatbib,usegraphicx]{mn2e}
\usepackage{graphicx}
\usepackage[latin1]{inputenc}
\usepackage{color}
\usepackage{times}
\usepackage{natbib}
\usepackage{setspace}
\usepackage{url}
\newif\ifAMStwofonts
\AMStwofontstrue
\definecolor{red}{rgb}{1,0.,0.}

\def\mincir{\lower.5ex\hbox{$\; \buildrel < \over \sim \;$}}
\def\magcir{\lower.5ex\hbox{$\; \buildrel > \over \sim \;$}}
\newcommand{\lya}{Lyman-$\alpha$\ }
\newcommand{\lyb}{Lyman-$\beta$\ }
\newcommand{\fesc}{f$_{\rm esc, q}$}
\newcommand{\fescg}{f$_{\rm esc, g}$}

\title[QSOs contribution to the UVB] {The Spectral Slope and Escape Fraction of Bright Quasars at $z \sim 3.8$:
the Contribution to the Cosmic UV Background}

\author[Cristiani et al.]{
  \parbox[t]{\textwidth}{ Stefano Cristiani$^{1,2}$,
  Luisa Maria Serrano$^{3,1}$,
  Fabio Fontanot$^{1}$,
  Rajesh R. Koothrappali$^{1}$,
  Eros Vanzella$^{4}$,
  Pierluigi Monaco$^{3}$
  }
    \vspace*{6pt}\\
    $^1$ INAF-Osservatorio Astronomico di Trieste, Via Tiepolo 11, I-34143 Trieste, Italy \\
    $^2$ INFN-National Institute for Nuclear Physics, via Valerio 2, I-34127 Trieste, Italy \\
    $^3$ Dipartimento di Fisica, Sezione di Astronomia, University of Trieste, via G.B. Tiepolo 11,
I-34143, Trieste, Italy\\
    $^4$ INAF-Osservatorio Astronomico di Bologna, via Ranzani 1, I-40127, Bologna, Italy\\
    email: cristiani@oats.inaf.it}

\begin{document}
\date{Accepted ...  Received Apr 1, 2016}

\maketitle

\begin{abstract}
We use a sample of 1669 QSOs ($r<20.15$, $3.6<z<4.0$) from the BOSS
survey to study the intrinsic shape of their continuum and the Lyman
continuum photon escape fraction (\fesc), estimated as the ratio
between the observed flux and the expected intrinsic flux (corrected
for the intergalactic medium absorption) in the wavelength range
865-885 \AA\ rest-frame.  Modelling the intrinsic QSO continuum shape
with a power-law, $F_{\lambda}\propto\lambda^{-\gamma}$, we find a
median $\gamma=1.30$ (with a dispersion of $0.38$, no dependence on
the redshift and a mild intrinsic luminosity dependence) and a mean
\fesc$=0.75$ (independent of the QSO luminosity and/or redshift).  The
\fesc\ distribution shows a peak around zero and a long tail of higher
values, with a resulting dispersion of $0.7$.  If we assume for the
QSO continuum a double power-law shape (also compatible with the data)
with a break located at $\lambda_{\rm br}=1000$ \AA \ and a softening
$\Delta\gamma=0.72 $ at wavelengths shorter than $\lambda_{\rm br}$,
the mean \fesc\ rises to $=0.82$.

Combining our $\gamma$ and \fesc\ estimates with the observed
evolution of the AGN luminosity function (LF) we compute the AGN
contribution to the UV ionizing background (UVB) as a function of
redshift.  AGN brighter than one tenth of the characteristic
luminosity of the LF are able to produce most of it up $z\sim 3$, if
the present sample is representative of their properties.  At higher
redshifts a contribution of the galaxy population is required.
Assuming an escape fraction of Lyman continuum photons from galaxies
between $5.5$ and $7.6\%$, independent of the galaxy luminosity and/or
redshift, a remarkably good fit to the observational UVB data up to
$z\sim 6$ is obtained.  At lower redshift the extrapolation of our
empirical estimate agrees well with recent UVB observations,
dispelling the so-called Photon Underproduction Crisis.
\end{abstract}

\begin{keywords}
  cosmology: observation - early Universe - quasars: general -
  galaxies: active - galaxies: evolution
\end{keywords}
\begin{figure}
  \centerline{ \includegraphics[angle=0, width=10cm]{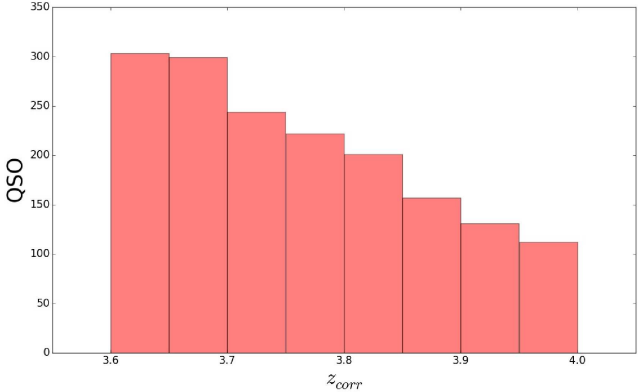} }
  \caption{Redshift distribution of the QSOs in the present sample
    shown in bins of $\Delta z = 0.05$.
$z_{corr}$ on the x-axis indicates the redshift estimated according to
the procedure described in Sect.~\ref{sec:datasample}
\label{fig:zhisto}
}
\end{figure}
\section{Introduction}
\label{sec:intro}
After more than thirty-five years \citep{Sargent80} the issue of the
sources driving the reionization of the hydrogen in the Universe and
keeping it ionized afterward does not appear to be settled.
It is commonplace that galaxies should be  able to produce the bulk of the UV
emissivity at high redshift (see, for example, \citeauthor{Robertson15}
\citeyear{Robertson15}),
but the AGN population is also proposed as a relevant or dominant
contributor (\citeauthor{Giallongo15} \citeyear{Giallongo15},
\citeauthor{Madau15} \citeyear{Madau15}, see also
\citeauthor{Fontanot12a} \citeyear{Fontanot12a} and
\citeauthor{Haardt15} \citeyear{Haardt15}, for different views).

A direct measurement of the 1-4 Ryd photons
escaping the various types of sources is unpractical at $z \magcir
4.5$, due to the reduced mean free path of these photons in the
intergalactic medium (IGM). 
At lower redshift direct observations of galaxies, 
after accounting for the statistical contamination of interlopers, 
have in general provided upper
limits in the fraction of ionizing photons, produced by young stars,
that are able to escape to the IGM (\fescg, see \citeauthor{Vanzella12}  \citeyear{Vanzella12}).
These limits tend to be significantly lower than the $ \sim 20\%$
required at $z\simeq 7$ to re-ionize the Universe with galaxies only
\citep{Bouwens11, HaardtMadau12}, and an increasing \fescg\ with decreasing luminosity
(possibly with a steep faint-end of the LF) has been
invoked to circumvent this shortcoming \citep{Fontanot14}.
The corresponding \fesc\ for QSOs is typically assumed to be about
$100\%$.

In this paper we aim to obtain a precise measurement of the
QSO contribution to the cosmic UV background in the
range $3.6<z<4.0$, where the QSO LF is well determined and the IGM
transmission not too low. Our strategy is 
first to estimate the intrinsic QSO continuum shape
up to 4 Ryd (Sect.~\ref{sect:SED}),
then we compare the fitted SED with the observed flux (corrected for
the effect of the IGM absorption with the model of \citet{Inoue14}),
and finally we compute the fraction of UV photons below the Lyman Limit
escaping to the IGM (\fesc, Sect.~\ref{sect:Fesc}).
We take advantage of the large samples of QSOs that can be extracted from the Sloan
Digital Sky Survey (SDSS) to investigate
possible correlations with the luminosity and redshift.
Finally, we combine this information with the knowledge of the QSO
LF to synthesize the global production of
ionizing photons from QSOs at various redshifts and compare it with
measurements of the UVB obtained from observations of the IGM
(Sect.~\ref{sect:UVB}).
In this way it is possible to assess how much room is left/needed for
the contribution of galaxies at the various cosmic epochs and where
preferably to look for it.
\section{Data Sample}
\label{sec:datasample}
The Baryon Oscillation Spectroscopic Survey (BOSS, \citet{Dawson13})
provides a large database of quasar spectra.
The quasar target selection used in BOSS is summarized in \citet{Ross12},
and combines various targeting methods described in \citet{Yeche10, Kirkpatrick11},
and \citet{Bovy11}.

We have extracted from the eleventh Data Release (DR11)
the quasars in the redshift range $3.6 < z \le 4.0$
with magnitudes brighter than $r = 20.15$.
The lower limit in the redshift interval is due to a known selection effet
in the BOSS survey outlined by \citet{Prochaska09a}:
the QSOs found in the range $3 < z < 3.6$ 
are selected with a bias against having $ (u-g)<1.5$, which
translates into a tendency to select sightlines with strong Lyman limit
absorption. On the other hand the analysis by \citet{Prochaska09a} shows that
beyond $z_{\rm em} = 3.6$ very few QSOs are predicted to have
such a blue $(u-g)<1.5$ color, removing the possibility of a bias.
We are therefore confident that the sample used can be considered
statistically complete and representative of the bright ($M_{\rm V}
\mincir -27.5$) QSO population. In particular, for the discussion to
follow, we note that BAL objects are included in the present sample.
The upper limit in the redshift range of the present sample, $z=4.0$,
is due to the requirement to have the observed spectra reaching the
rest-frame wavelength of $2000$ \AA\ in order to have a sufficiently
extended domain to estimate the intrinsic QSO continuum shape.

The spectral energy distribution (SED) of each quasar has been adjusted
using a linear multiplicative slope (in magnitude) in order to match
the $g$, $r$, $i$, $z$ magnitudes from the SDSS photometric catalog
and then corrected for galactic extinction according to the maps of
\citet{Schlafly11} and the average Milky Way extinction curve of
\citet{Cardelli89}. 

All the spectra have been visually inspected and
their systemic redshifts calculated.
We have adopted the offsets of $-310$ km/s and
$+177$ km/s, respectively assigned to the CIV 1549
and SiIV 1398 lines \citep{Tytler92} to derive the systemic redshift. 
A small fraction ($ \mincir 1\%$) of spectra showing problems
in terms of the observed S/N ratio have been excluded
from the subsequent analysis.
The resulting sample consists of 1669 objects and the associated 
redshift distribution is shown in Fig.~\ref{fig:zhisto}. 
The number of QSOs declines from about 300 per $\Delta z =0.05$ bin
in the interval $3.6 < z < 3.7$ to about 100 at $z\simeq 4$, following
the general trend of the BOSS Survey. 
The distribution of the recomputed redshifts shows a small systematic
difference with respect to the SDSS data, $<{\Delta z}> = 
<{z_{corr}-z_{SDSS}}> = -0.008$, with a dispersion of $0.012$.
\begin{figure*}
  \centerline{ \includegraphics[angle=270, width=18cm]{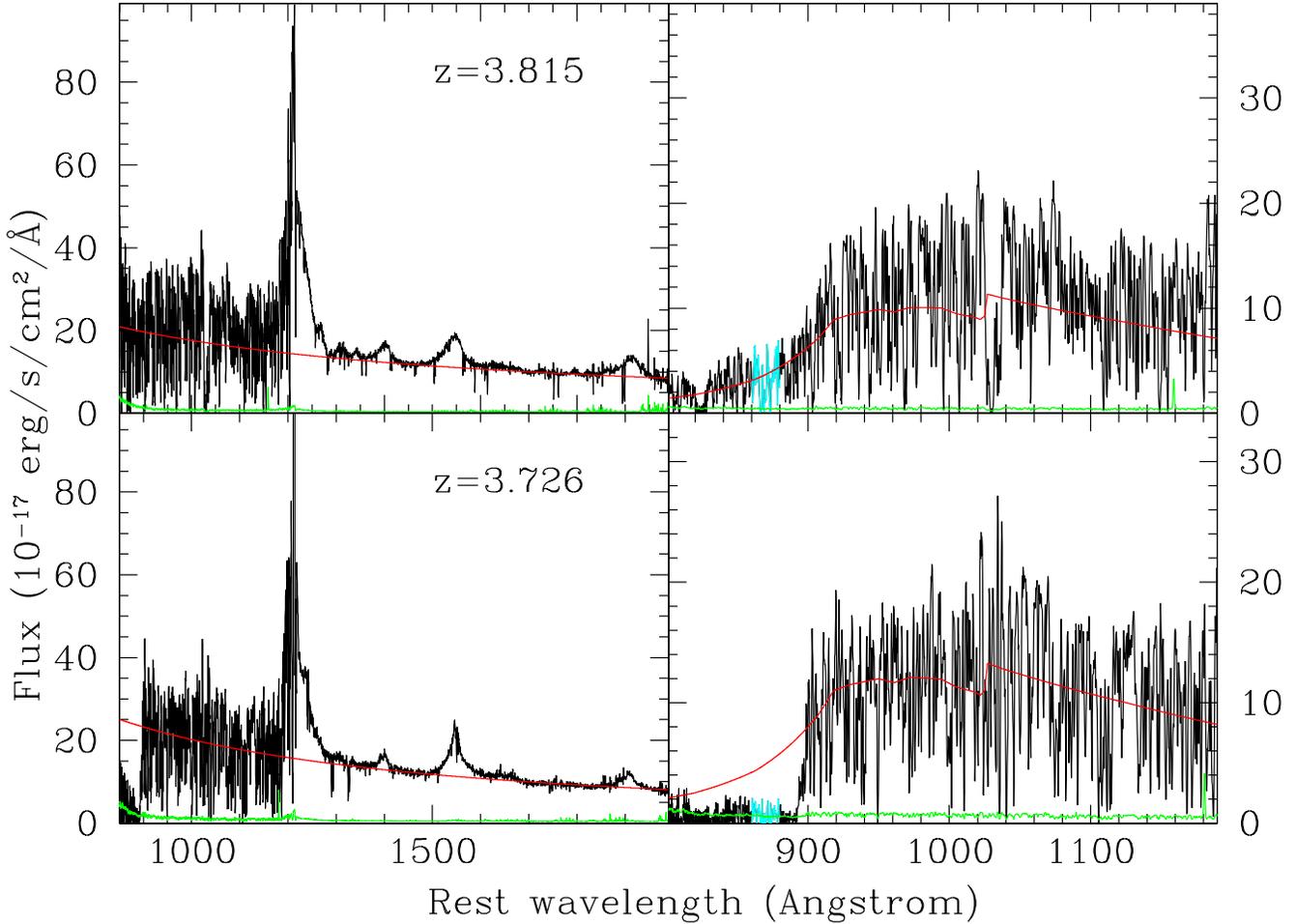} }
  \caption{Two illustrative cases for the estimate of the continuum
power-law and the escape fraction.
The upper panels show a QSO of redshift 3.815 with a $f_{\rm esc, q} \sim 1$,
the lower panels a QSO of redshift 3.726 with a $f_{\rm esc, q} \sim 0$.
{\it Left column}: the two spectra are plotted in black. Note that
the region blueward of the \lya has been divided by the average
transmission of the IGM estimated (see Sect.~\ref{sect:Fesc})
according to \citet{Inoue14} in order to show where the average
continuum should be located.
The green line shows the uncertainty of
the flux. The red line shows the fitted power-law continuum.
{\it Right column}: the two observed spectra are plotted in black,
the expected position of the power-law continuum multiplied by
the average IGM transmission is in red and
the uncertainty of the observed flux in green.
The cyan portion of the spectrum is the region where the escape
fraction of the UV photons produced by the QSO has been estimated.
\label{fig:spectra}
}
\end{figure*}
\section{Estimate of the QSO Spectral  Energy Distribution}
\label{sect:SED}
In order to estimate the QSO production of UV ionizing photons 
it is necessary to model their intrinsic spectral shape.
Customarily this is achieved by fitting a power-law, 
$F_{\lambda} \propto \lambda^{-\gamma}$, 
in the region redward of the \lya emission, selecting windows free of
emission lines and extrapolating it in the region blueward of the \lya
(e.g. Fig.~\ref{fig:spectra}).
Previous works
by \citet{Zheng97, Telfer02, Shull12, Stevans14}
 at relatively low redshift ($z \mincir 1.5$), 
where the IGM absorption is minimized,  
have identified a break in the spectral distribution with a
softening of the slope at wavelengths shorter than $\lambda_{\rm br}
\sim 1000$ \AA.

In the following we have chosen to fit the continuum spectrum of each
quasar both with a single and with a broken power-law.
In both cases five windows, listed in Tab.~\ref{tab:windows},
have been used for the fit as emission-line-free regions.
In the case of the broken power-law fit we have imposed 
a flattening in the spectral slope blueward of $1000$ \AA\
of $\Delta \gamma = 0.72$ with respect to the power-law at longer wavelengths,
as reported by \citet{Stevans14}.
Pixels affected by absorption lines have been iteratively rejected
on the basis of a three sigma k-clipping.
We have checked that the results are not sensitive to the particular
choice of the windows.
\begin{table}
 \centering
  \caption{Regions used for the continuum fitting}
  \begin{tabular}{@{}crc@{}}
  \hline
   Region    & start &  end \\
             & \multicolumn{2}{c}{rest-frame wavelength} \\
             & \multicolumn{2}{c}{(\AA)} \\
 \hline
1  & 1990 & 2020 \\
2  & 1690 & 1700 \\
3  & 1440 & 1465 \\
4  & 1322 & 1329 \\
5  & 1284 & 1291 \\
\hline
\end{tabular}
\label{tab:windows}
\end{table}

\begin{figure}
\centerline{ \includegraphics[angle=270, width=9.1cm]{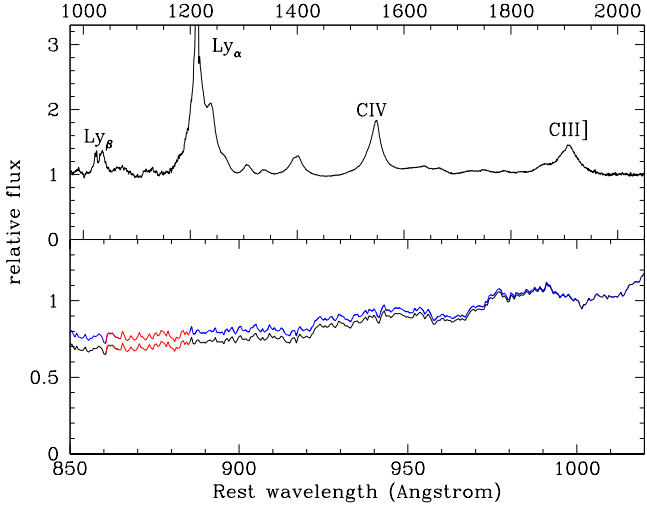}}
 \caption{Stacked spectrum of the 1669 BOSS QSOs, 
after dividing each QSO spectrum by the average
IGM transmission and by the estimated individual power-law continuum.
Upper panel: $1000-2000$ \AA\ rest-frame wavelength range.
Lower panel: $850-1020$ \AA\ wavelength range.
In the lower panel the lower black line corresponds to a single
power-law continuum, while the upper blue line to the broken power-law
fitting (see text).
The region $865-885$ \AA\ restframe, where the escape fraction has been
measured (see Sect.~\ref{sect:Fesc}), is shown in red.
}
\label{fig:stacking}
\end{figure}

As a check of the goodness of the assumptions, we have stacked all the
spectra after dividing them by the continuum slope and by the expected
mean transmission of the IGM according 
to the computation by \citet{Inoue14}.
For the IGM transmission
we have used the numerical tables kindly provided by the authors,
which are slightly more accurate than the analytical approximation,
especially in the region between the \lyb and the Lyman limit.
The result is shown in Fig.~\ref{fig:stacking}.
The continuum normalized average flux, corrected by the IGM absorption, 
in a true-continuum window of the \lya forest 
($1080<\lambda<1120$, see Shull et al. 2012 and Stevans et al. 2014) 
turns out to be $1.01 \pm 0.04$ (see Fig.~\ref{fig:stacking}).
It is also remarkable to see the correspondence between
the emission bumps observed in the Lyman forest by us and by \citet{Shull12}
and \citet{Stevans14},
in particular around $1125$\AA\ (Fe III) and $1070$ \AA (NII, He II).

\begin{figure}
\centerline{ \includegraphics[angle=0, width=10cm]{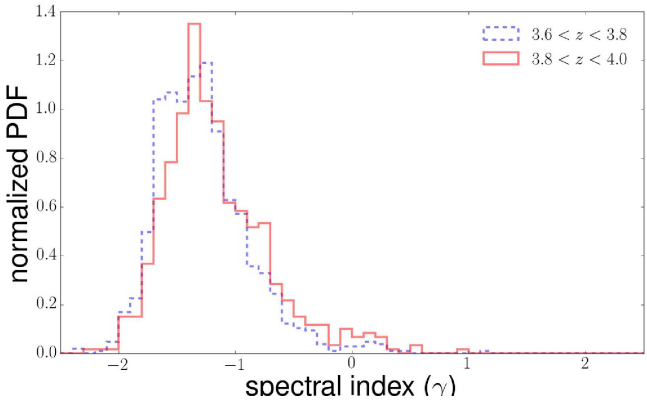} }
 \caption{Normalized probability distributions of the spectral index
   $\gamma$ for QSOs in the redshift interval $3.6 < z \leq 3.8$ (blue
   dashed line) and
$3.8 < z \leq 4.0$ (red continuous line).
}
\label{fig:PDFgamma}
\end{figure}

\begin{figure}
\centerline{ \includegraphics[angle=0, width=10cm]{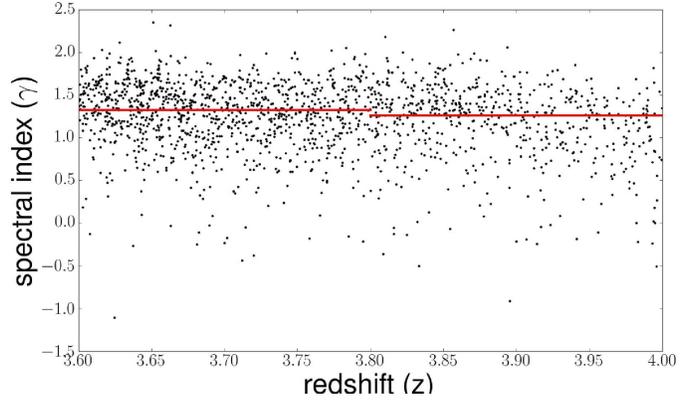} }
  \caption{Distribution of the spectral indices of the continuum power-law
of the QSO spectra as a function of the redshift.
The median values in the intervals
 $3.6< z \leq 3.8$, $3.8< z \leq 4.0$,
are shown as continuous red segments.
}
\label{fig:gammas}
\end{figure}

We have then analyzed the ensemble properties of the quasars
in our sample.
The median (mean) spectral index of the population for the single power-law case
(and for the region with $\lambda > 1000$ \AA for the broken power-law)
is $\gamma = 1.30$ ($1.24$), with a dispersion of $0.38$, computed as half of the
difference between the 84.13 and 15.87 percentiles (Fig.~\ref{fig:gammas}).
A KS test on the two samples above and below redshift $z=3.8$ does not show any significant
difference in the two distributions (see also
Fig.~\ref{fig:PDFgamma} and \ref{fig:gammas}).
We have checked that selecting QSOs with $3.4<z<3.6$ (and $r<20.15$) we would obtain a
value of $\gamma$ significantly lower than the one we measure in the
range $3.6 < z \le 4.0$, confirming the above mentioned bias
found by \citet{Prochaska09a}.
The dependence of the spectral index on the SDSS $r$ magnitude has
also been analyzed by splitting the sample in two halves:
$r \leq 19.69$ and $r > 19.69$.
The corresponding median spectral indices turn out to be $\gamma =
1.36$ and $\gamma = 1.22$, respectively, with fainter objects
generally characterized by ``redder'' spectral indices.
A KS test rejects with a high significance
the hypothesis that the two subsamples have the same distribution
function.
We interpret this effect as a property of the SEDs of the QSOs
analyzed, rather than a bias introduced by the fitting procedure,
since a corresponding difference is present in the measured colors of
the QSOs: the average $(r-i)$ is $0.115$ for objects brighter than
$r=19.69$ and $0.138$ for the fainter ones.

The median (mean) value of the spectral index for the full sample,
$\gamma = 1.30$ ($1.24$), can be compared with the results based 
on Hubble Space Telescope (HST) COS data at $z<1.5$
by \citet{Shull12}, who find $\gamma = 1.32 \pm 0.14$ and 
\citet{Stevans14} who measure $\gamma = 1.17 \pm 0.09$.
\citet{Telfer02} with a similar approach and wavelength windows to
ours, at $<z> = 1.17$ find a $\gamma = 1.31 \pm 0.06$ with HST 
FOS, GHRS and STIS data.
\section{The average escape fraction of the  QSO population at $3.6  < z \leq 4.0$}
\label{sect:Fesc}
The fraction of the UV ionizing photons produced by each QSO
leaking to the IGM 
has been estimated by dividing the observed average
flux in the region $865-885$ \AA\ rest frame 
by the expected average flux in the same region, as estimated either
with the single or with the broken power-law, convolved with the
average transmission of the IGM at the given redshift \citep{Inoue14}.
The interval $865-885$ \AA\ has been chosen 
since its is expected to be a ``true continuum'' window (see Fig.6 in
Shull et al.~2012 and Fig.5 in Stevans et al.~2014).
Besides, it represents a convenient compromise: 
on the one hand at wavelengths close to the Lyman edge the measurement
can be affected by errors in the determination of the emission redshift
of the QSO,
on the other hand the IGM transmission is progressively decreasing at shorter
and shorter wavelengths with a consequent increase of the measurement uncertainty.
The resulting value of the \fesc\ 
has been checked to be largely independent of the specific choice
of the limits of the interval.

The estimated \fesc\ is an {\it effective} escape fraction, i.e. is
expected to include the escape fraction of the UV photons from the QSO
host galaxy and all the extra absorption due to clustered neutral
hydrogen in the vicinity of the QSO that is not accounted for in the
model of \citet{Inoue14} which applies to the average, intervening IGM.

\begin{figure}
\centerline{ \includegraphics[angle=0, width=10cm]{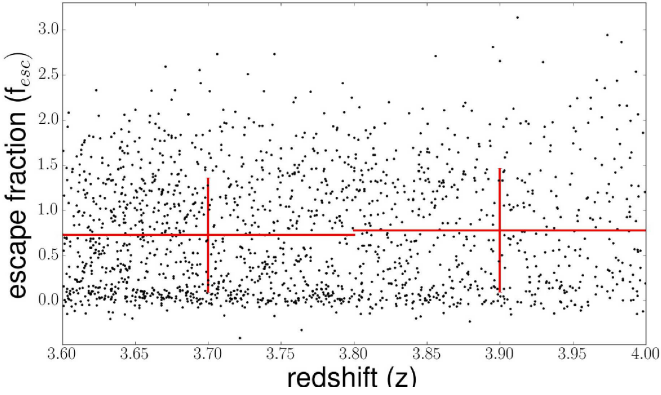} }
 \caption{Escape fraction measured in the
QSO spectra as a function of the redshift.
The mean values in the intervals
 $3.6< z \leq 3.8$ and $3.8< z \leq 4.0$
are shown as continuous red segments, with the
dispersion estimated as half of the
difference between the 84.15 and 15.87 percentiles.
}
\label{fig:fescz}
\end{figure}

\begin{figure}
\centerline{ \includegraphics[angle=0, width=10cm]{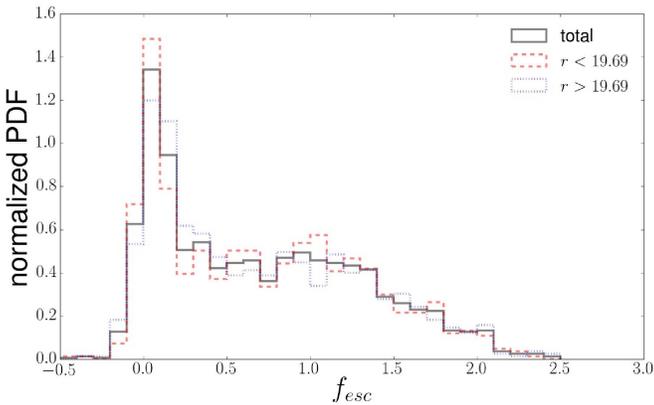}}
 \caption{Normalized probability distributions of the escape fraction \fesc\
   for QSOs in the redshift interval $3.6 < z \leq 4.0$. The black
   continuous line shows the full sample. The red dashed line
   corresponds to objects with $r \leq 19.69$, while the blue dashed
   line refers to objects with $r > 19.69$.
}
\label{fig:PDFfesc}
\end{figure}

The average escape fraction in the redshift interval $3.6<z<4.0$,
measured on the ensemble of 1669 objects of our sample turns out to be
\fesc $=0.75$ and \fesc $=0.82$ in the case of the single and broken
power-law respectively.
As shown in Fig.~\ref{fig:fescz} and \ref{fig:PDFfesc}, a rather large dispersion is
observed, $0.7$, computed as half of the
difference between the $84.13$ and $15.87$ percentiles, in a kind of
bimodal distribution with a narrower peak around the value zero and a
larger dispersion around the value $1$.
In each object \fesc\ is computed by comparing the observed flux
shortward of the Lyman edge and the expected flux on the basis of an
average correction for the IGM absorption. It is not surprising therefore
that for some objects our measured \fesc\ turns out to be larger
than one - besides the measurement errors - 
due to lines of sight with an actual transmission 
larger than the average estimate from the \citet{Inoue14} computation.

Fig.~\ref{fig:fescz} shows the escape fraction measured in the QSO
spectra as a function of the redshift.
In the intervals $3.6 < z \leq 3.8$ and $3.8 < z \leq 4.0$ the result is
similar: \fesc $= 0.73$ and \fesc $= 0.78$, respectively for the
single power-law, \fesc $= 0.80$ and \fesc $= 0.85$, respectively, for the
broken power-law.  

Splitting the sample in two halves, brighter and fainter than $r=19.69$
does not show any significant difference in the mean \fesc\ 
and a KS test cannot reject the hypothesis that the parent
population is the same for the two groups, both in the case of a 
single and of a broken power-law.

No significant correlation of \fesc\ is therefore found as a function neither of the
redshift (see Fig.~\ref{fig:fescz}), nor of the magnitude
(Fig.~\ref{fig:PDFfesc}).

We have also checked that no dependence of \fesc\ is present as a
function of the spectral index $\gamma$: dividing sample in two
halves, the average \fesc\ of QSOs with $\gamma \ge 1.296$ is $0.77$,
while \fesc\ $= 0.73$ for QSOs with  $\gamma < 1.366$, for the single
power-law,
$0.84$ and $0.80$ for the broken power-law.
\section{Synthesis of the Ionizing Background}
\label{sect:UVB}
\begin{figure}
  \centerline{ \includegraphics[width=9cm]{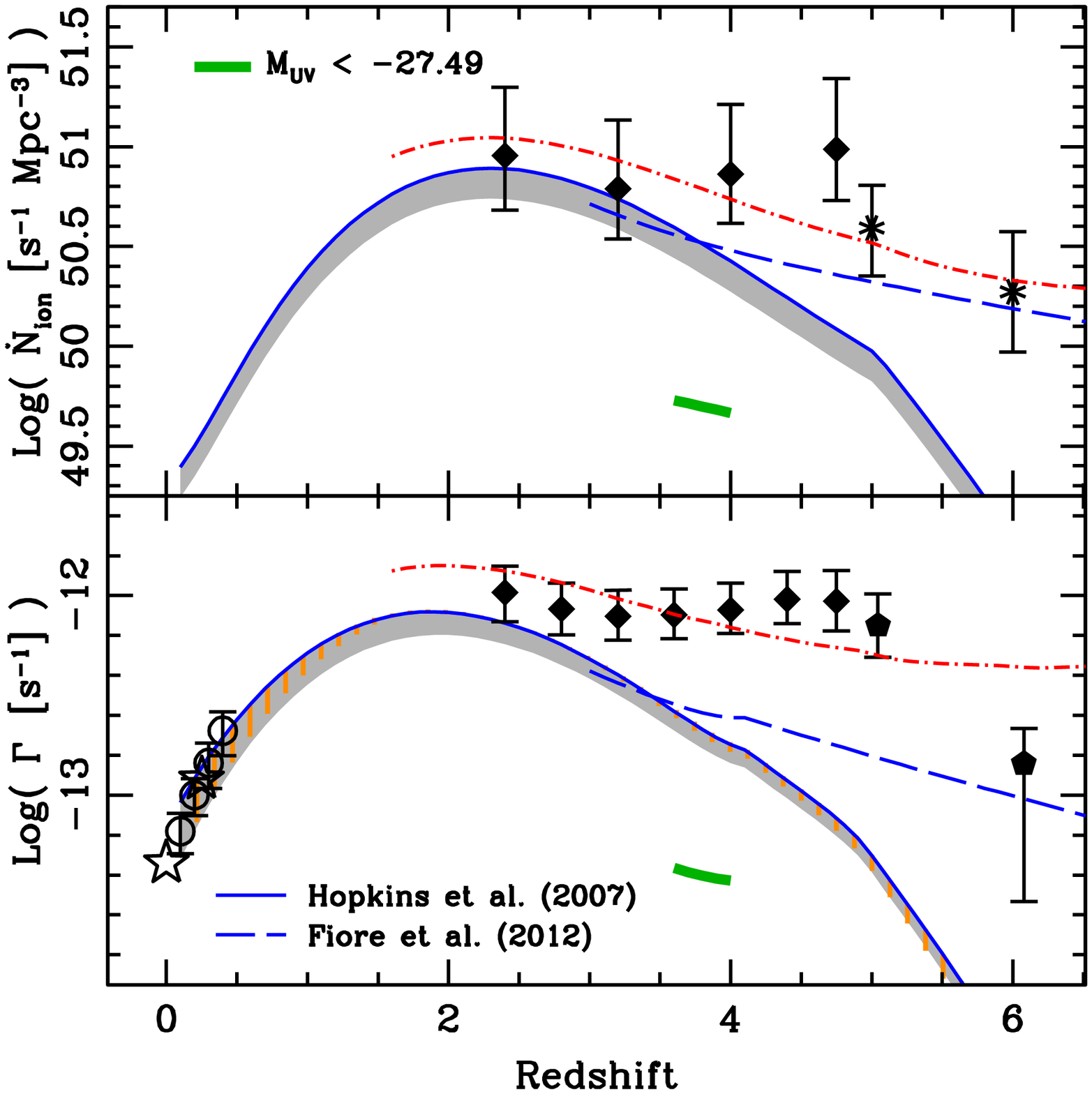} }
  \caption{{\it Upper panel}: predicted photon volume
      emissivity. Observed data from \citet{Wyithe11} (asterisks) and
    \citet{Becker13} (diamonds). {\it Lower Panel}: predicted hydrogen
    photoionization rate. Observations from
    \citet[diamonds]{Becker13},
    \citet[pentagons]{Calverley11}, \citet[stars]{Shull15} and
      \citet[empty circles]{Gaikwad16}. In both panels, solid and
    dashed lines (in blue) represent the predictions corresponding to
    the AGN-LF from~\citet{Hopkins07} and~\citet{Fiore12},
    respectively, integrated up to $0.1~ L_\star$ assuming a single
    power-law quasar SED. The grey area extends down to the double
    power-law results, to show the deriving systematic uncertainty,
    while the hatched orange area represents the uncertainty
    relative to the shape of the assumed column density
    distribution (see text). The short thick segments (in green) in the
    redshift range $3.6<z<4$, show the contribution of QSOs brighter
    than $M_{\rm UV} \sim -27.49$ roughly corresponding to the
    absolute magnitude limit in the present sample, assuming the
    \citet{Hopkins07} bolometric LF.  The dot-dashed red lines show
    the total UV background and photoionization rate adding to the
    blue solid line a contribution of the galaxy population estimated
    assuming an $f_{\rm esc, g} = 5.5 \%$ (see text for more details).
  }\label{fig:ibck}
\end{figure}

We use the results of the previous section to estimate the QSO
contribution to the observed photon volume emissivity
(Fig.~\ref{fig:ibck}, upper panel) and photoionization rate (lower
panel), adopting the same formalism as in \citet{Fontanot14}.

We consider functional forms for the AGN LF $\Phi(L,z)$ as a function
of luminosity and redshift and we use them to compute the rate of
emitted ionizing photons per unit comoving volume as a function of the
redshift:

\begin{equation}
\dot{N}_{\rm ion}(z) = \int_{\nu_H}^{\nu_{\rm up}}  \frac{\rho_\nu}{h_p \nu} d\nu
\end{equation}

\begin{equation}\label{eq:lfint}
\rho_\nu = \int_{L_{\rm min}}^\infty f_{\rm esc}(L,z) \, \Phi(L,z) \, L_\nu(L) \, dL
\end{equation}

\noindent
where $\nu_H$ is the frequency corresponding to $912$ \AA\ and
$\nu_{\rm up} = 4 \nu_H$ (i.e. we consider that more energetic photons
will be mainly absorbed by He {\sc II} atoms), while $\rho_\nu$ is the
monochromatic comoving luminosity density brighter than $L_{\rm
  min}$. The redshift evolution of the corresponding photoionization
rate $\Gamma$ is computed solving the following equations (see
e.g. \citet{HaardtMadau12} and references therein):

\begin{equation}
\Gamma(z) = 4 \pi ~ \int_{\nu_H}^{\nu_{\rm up}} \frac{J(\nu,z)}{h_p \nu} \sigma_{HI}(\nu) d\nu
\end{equation}

\noindent
where $\sigma_{HI}(\nu)$ is the absorbing cross-section for neutral
hydrogen and $J(\nu,z)$ is the background intensity:

\begin{equation} \label{eq:Jnu}
J(\nu,z) = c/4 \pi \int_{z}^{\infty} \epsilon_{\nu_1}(z_1)
e^{-\tau_{\rm e}} \frac{(1+z)^3}{(1+z_1)^3} \, |\frac{dt}{dz_1}| \,
dz_1
\end{equation}

\noindent
where $\nu_1 = \nu \frac{1+z_1}{1+z}$, $\epsilon_\nu(z)$ represents
the proper volume emissivity (equivalent to $\rho_\nu$ in the
  comoving frame) and $\tau_{\rm e}(\nu,z,z_1)$ the effective opacity
between $z$ and $z_1$:

\begin{equation}
\tau_{\rm e}(\nu,z,z_1) = \int_{z}^{z_1} dz_2 \int_{0}^\infty  dN_{HI} f(N_{HI},z_2) (1-e^{-\tau_{\rm c}(\nu_2)})
\end{equation}

\noindent
where $\tau_{\rm c}$ is the continuum optical depth through an
individual absorber at frequency $\nu_2 = \nu \frac{(1+z_2)}{(1+z)}$
and $f(N_{HI},z)$ is the bivariate distribution of absorbers. For
  the latter quantity, we consider different functional forms
  available in the literature, namely those proposed by
  \citet{HaardtMadau12}, \citet{Becker13} and
  \citet{Inoue14}. In the following, we adopt
  \citet{Becker13} as a reference, because we want to compare our
  predictions for the photon volume emissivity and photoionization
  rate in particular with their dataset, which covers a redshift range
  encompassing our sample.
\begin{figure}
  \centerline{ \includegraphics[width=9cm]{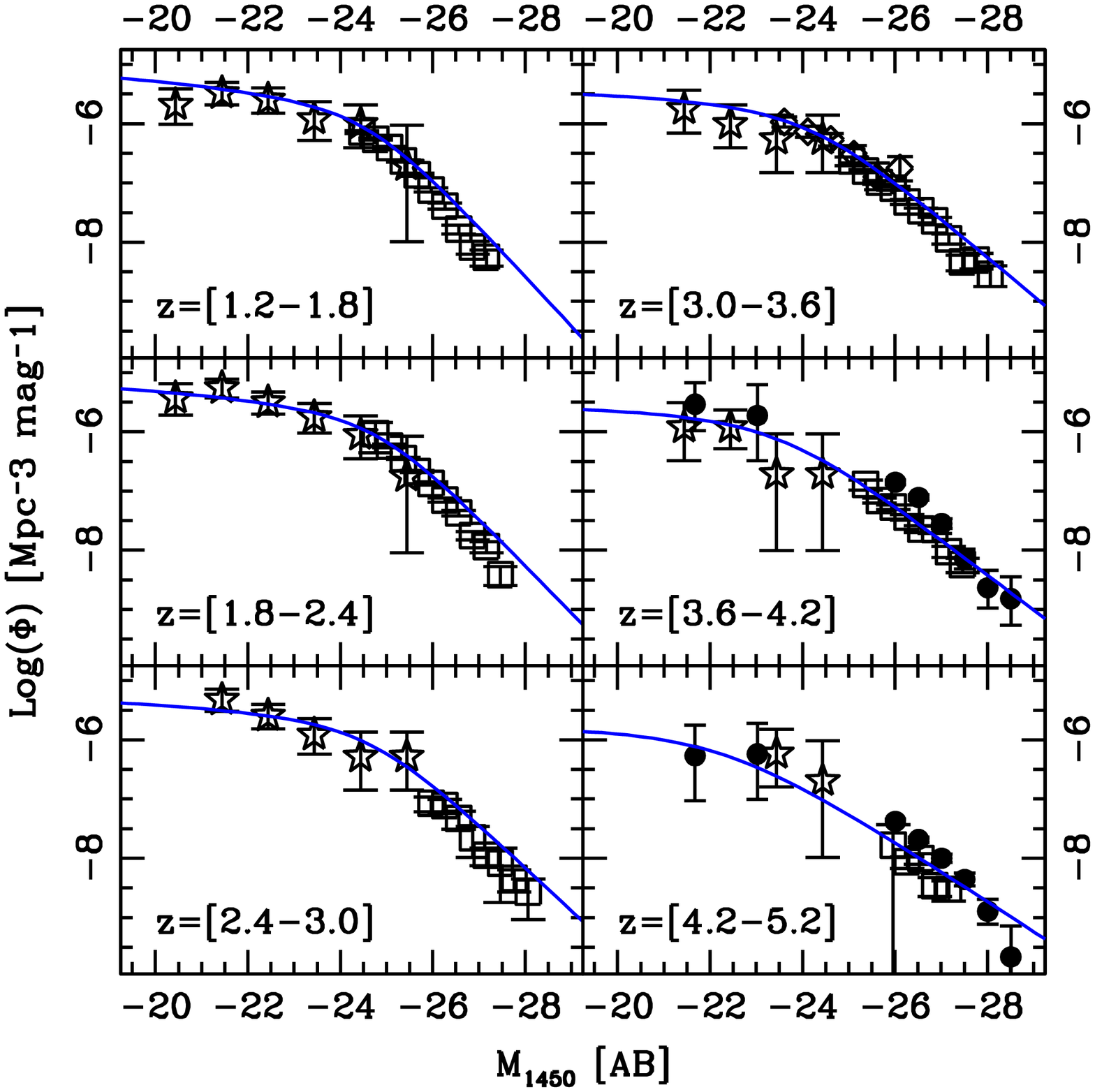} }
  \caption{QSO Luminosity function at 1450 \AA. Solid blue lines refer
    to the analytical fits from \citet{Hopkins07} and are compared to
    observational estimates from \citet[stars]{Wolf03} , \citet[open
      squares]{Richards06}, \citet[filled circles]{Fontanot07} and
    \citet[open diamonds]{Siana08}.}
\label{fig:lf}
\end{figure}
We consider two different estimates for the AGN-LF, namely the
luminosity function at 145 nm (see Fig.~\ref{fig:lf}) defined in the
framework of the \citet{Hopkins07} bolometric LF and the Hard X-ray LF
from \citet{Fiore12}.  We use the resulting space densities in
Eq.~\ref{eq:lfint} and \ref{eq:Jnu}, we then integrate Eq.1 and 3
using the median spectral index from Sect.~\ref{sect:SED} and using
the corresponding $L_{145}$ as a normalization.  In
Fig.~\ref{fig:ibck}, the solid line represents predictions
corresponding to the \citet{Hopkins07} 145 nm LF at $z < 4$ (and its
extrapolation at higher redshifts), while dashed line refers to the
\citet{Fiore12} LF ($z > 3.5$), assuming a single power-law SED. We
adopt as $L_{\rm min}$ one tenth of the characteristic luminosity of
the LF (i.e. $L_{\rm min} = 0.1 ~ L_\star$).  \citet{Cowie09} have
shown, in fact, that most of the ionizing flux is produced by
broad-line QSOs straddling the break luminosity.  Although our
formulation allows for a luminosity and redshift dependent escape
fraction, we assume a fixed \fesc\ $=0.75$, consistently with the
results in Sect.\ref{sect:Fesc}.

In Fig.~\ref{fig:ibck}, we use hatched and grey areas to highlight
  the effect of two of the main uncertainties involved in the estimate
  of the photon volume emissivity and photoionization rate. In
  particular, the hatched orange area represents the variation 
  corresponding to different functional forms for the column density
  distribution \citep{HaardtMadau12, Becker13, Inoue14}, 
  while the grey area refers to the difference between the single
  and the broken power-law assumption for the AGN spectral shape. To the
zeroth order, adopting a single power-law with a 0.75 \fesc\ or a
broken power-law with a 0.82 \fesc\ is degenerate from the point of
view of the UV background: the assumption of the SED type is
compensated by the resulting \fesc\ and the same flux is predicted at
the Lyman Limit.  The difference between the two predictions arises
from the extrapolation of the flux up to 4 Ryd with different slopes.

Our estimates are then compared with a collection of observational
results for the photon volume emissivity \citep{Wyithe11,
  Becker13} and photoionization rate \citep{Calverley11, Adams11,
  Becker13, Shull15, Gaikwad16}. It is worth stressing that our
  estimates do not exactly correspond to the predictions of the
  \citet{HaardtMadau12} model. The main difference lies in the 
  assumption by \citet{HaardtMadau12} of a QSO emissivity based on the 
  \citet{Hopkins07} LF with a contribution of relatively bright ($M_B<-27$) QSOs
  only. 
  Here, we are considering objects down to $0.1 ~ L_\star$, which implies
  a fainter (and variable with redshift) limiting magnitude.

Our predictions are consistent with a number of observational
constraints, and in particular with the data at $2<z \mincir 3$ from
\citet{Becker13}: this suggests both that sources brighter than $0.1~
L_\star$ account for the observed ionizing photons and, conversely,
that objects fainter than $0.1 ~ L_\star$ should provide a negligible
contribution to the ionizing photon budget.  It will be therefore of
interest to test with future observations the \fesc\ for
low-luminosity QSOs to check whether smaller values with respect to
the present sample are measured. There is already an indication from
the observations of \citet{Cowie09} that this is indeed the case.  The
thick green segments in Fig.~\ref{fig:ibck} spanning the redshift
range of the present sample represent the integration of the
\citet{Hopkins07} LF up to $M_{\rm UV} \sim -27.49$, roughly
corresponding to the absolute magnitude limit in our QSO sample, in
the range $3.6 < z \mincir 4.0$. They lie $\sim 0.8$ dex below the
solid lines, highlighting that the QSOs in the present sample account
for less than one sixth of the full background and, again,
observations of fainter objects would be advisable in order to avoid
extrapolations. The prediction obtained with the luminosity function
by \citet{Fiore12} highlights the effect of the uncertainties in the
LF estimate and the need for a better determination of this
distribution at high-z.

We confirm that the QSO cannot dominate the ionizing photons
production at $z > 4$: in fact none of our predictions reproduces the
observational data, typically underestimating them, thus highlighting
the need for additional ionizing sources at these redshifts
(e.g. galaxies, dot-dashed red lines in Fig.~\ref{fig:ibck}). The
contribution from galaxies has been computed from the LF of Lyman
Break Galaxies \citep{Bouwens11} using eq. 1-5, assuming the
redshift-depedent spectral emissivity as in \citet{HaardtMadau12},
the column density distribution as in \citet{Becker13} and a
constant value for the \fescg\ (i.e. independent of either the
luminosity or the redshift). The corresponding LFs have been
integrated up to a limiting redshift-dependent faint magnitude
computed as in \citet[their Fig.~3]{Fontanot14}.

If we limit the analysis to the photon volume emissivity shown
in the upper panel of Fig.~\ref{fig:ibck} and fix the contribution of
the QSO population to the above determined {\it bona fide} amount
(shown by the blue solid line), then in the case of a single power-law
quasar SED the best fit to the observational data (with a $\chi^2 =
2.3$ for six points and one free parameter) turns out to be \fescg\ $
= 6.9^{+6.3}_{-4.2}\%$, with the confidence interval estimated for
$\Delta\chi^2 = 1$.  If we apply the same analysis to the
photoionization rate (lower panel of Fig.~\ref{fig:ibck}) we obtain a
best fit \fescg\ $ = 5.5^{+3.4}_{-1.2}\%$, with a $\chi^2 = 9.8$ for
nine points and one free parameter). 
In the case of a double power-law
quasar SED the best fit to the upper panel (with a $\chi^2 = 2.1$ for
six points and one free parameter) turns out to be \fescg\ $ =
7.6^{+6.7}_{-3.8}\%$, and for the lower panel we obtain a best fit
\fescg\ $ = 6.0^{+2.3}_{-1.3}\%$, with a $\chi^2 = 8.9$. It is
interesting to note that all these values are fully compatible with
the limits obtained by various authors with direct measurements
of the \fescg\ (e.g. \citet{Vanzella12, Bouwens15, Reddy16}).

Finally, at $z<1$ our estimates agree
with the most recent determinations for the HI photoionization rate 
by \citet{Shull15} and \citet{Gaikwad16}, based on HST-COS data
\citep{Danforth16}. 
Both groups find values of the photionization rate
significantly smaller than the results presented in
\citet{Kollmeier14}, giving origin to the so-called 
Photon Underproduction Crisis (PUC).
Our computation shows that
relatively bright QSOs at low redshift (i.e. brighter than
$0.1 ~ L_\star$) may account for the total photon budget required by
observations.
A similar result has been obtained by \citet{Khaire15}.

\section{Conclusions}
\label{sec:concl}
In this paper, we use a sample of 1669 QSOs with $r \leq 20.15$ in the
redshift range $3.6<z<4.0$, taken from the BOSS sample, to estimate
the contribution of type I QSOs to the UV background. Each spectrum in
the sample has been recalibrated to match the observed SDSS
photometry, and the corresponding systemic redshift has been recomputed,
taking into account the velocity shifts associated with the ${\rm SiIV}$ and
${\rm CIV}$ emission lines. 
For each QSO, we fit the intrinsic continuum spectrum,
by means of five windows relatively free of emission lines,
both with a single, $F_{\lambda} \propto \lambda^{-\gamma}$,  and with a broken power-law
with a break located at $\lambda_{\rm br}=1000$ \AA\ rest-frame.

In order to constrain the Lyman continuum photon escape fraction,
\fesc\, in our sample,
we consider the spectral
range $865-885$ \AA\ rest frame, close to the Lyman limit. 
We compute \fesc\ as the ratio between observed flux in this interval
and the flux expected on the basis of the intrinsic quasar continuum 
and the average attenuation due to the IGM.

Using our reference sample, we estimate a median $\gamma = 1.30$, with
a dispersion of $0.38$ in its distribution, and a mean \fesc\ $= 0.75$ 
in the case of the single power-law fit and $= 0.82$ for the broken power-law.

We do not find any evidence for a redshift dependence of both
quantities. 
$\gamma$ shows a small dependence on the $r$-mag, which is likely due to an intrinsic
effect, with the fainter sources having flatter continua ($\gamma =
1.36$ for QSOs brighter than $r=19.69$ and $\gamma=1.22$ for
$r>19.69$). 
The statistical distribution of \fesc\ is characterized by a kind of 
bimodality: this shape suggest an interpretation of \fesc\ 
as a probabilistic distribution, rather than a mean value,
with $\sim 25-18\%$ of the object characterized by a negligible escape
fraction and the rest with roughly clear lines of sight.
For comparison the percentage of BAL quasars in the BOSS survey
has been estimated to be around $10-14 \%$ \citep{Paris14, Allen11}. 
No dependence of \fesc\ with luminosity is present in our sample.

We have combined the observed evolution of the AGN/QSO-LF with
our measurement of the escape fraction to compute the expected rate of
emitting ionizing photons per unit comoving volume $\dot{N}_{\rm ion}$
and photoionization rate $\Gamma$, as a function of redshift. We show
that, given our mean values for \fesc, $L > 0.1 ~ L_\star(z)$ sources
are able to provide enough photons to reproduce the reionization
history in the redshift interval $2 < z \mincir 3$, while we confirm that at
$z \magcir 4$ additional sources of ionizing photons are required. However,
the details on the reionization history are affected by the
uncertainties in the QSO luminosity function evolution as estimated in
the optical and X-ray bands.

Overall, our results imply that, at $2<z<4$, the contribution to the
ionizing background of AGNs fainter than the LF characteristic
luminosity, $0.1 ~ L_{\star}$, should be negligible. 
Since our sample covers only magnitudes brighter
than $M_{\rm UV} \sim -27.49$, we also forecast that fainter QSOs (but
still brighter than $0.1 ~ L_\star$) should be characterized by an \fesc\ 
as large as those found in this work, in order for the QSO
population to account for the whole photon budget at the redshift of
interest. 

Our predictions are perfectly
compatible with the low redshift estimate of
\citet{Shull15, Gaikwad16}, suggesting that QSOs brighter than 
$0.1 ~ L_\star$ may account for the total photon budget at low redshift.

At $z>4$ a contribution to the UV background from the galaxy population
is needed. 
A good fit
from $z=2$ to $z=6$ of the data is obtained
assuming an escape fraction \fescg\ between $5.5$ and $7.6\%$
(depending on the assumptions on the quasar SED and the comparison
with the ionizing background or photoioniziation rate measurements), 
independent of the galaxy luminosity and/or redshift, 
added to the present determination of the QSO contribution.

On the basis of the present approach, future area of progress,
besides the obvious direct determination of \fescg$(L,z)$,
are linked to a better knowledge of
the QSO luminosity function,
the \fesc\ for fainter quasars (at least down to  $0.1 ~ L_\star$)
and its possible dependence on the redshift,
the intensity of the UVB, which in turn requires improved
    simulations of the IGM.
\section*{Acknowledgments}
We are grateful to A.Inoue, S. Cooper,
E. Giallongo, F.Haardt, L. Hofstadter, J.Japelj, I. P\^aris
and H. Wolowitz,
for providing unpublished material and
enlightening discussions. 
We acknowledge financial support from the
grants PRIN INAF 2010 ``From the dawn of galaxy formation'' and PRIN
MIUR 2012 ``The Intergalactic Medium as a probe of the growth of
cosmic structures''.
Funding for SDSS-III has been provided by the Alfred P. Sloan
Foundation, the Participating Institutions, the National Science
Foundation, and the U.S. Department of Energy Office of Science. The
SDSS-III web site is \url{http://www.sdss3.org/}.
SDSS-III is managed by the Astrophysical Research Consortium for the
Participating Institutions of the SDSS-III Collaboration including the
University of Arizona, the Brazilian Participation Group, Brookhaven
National Laboratory, University of Cambridge, Carnegie Mellon
University, University of Florida, the French Participation Group, the
German Participation Group, Harvard University, the Instituto de
Astrofisica de Canarias, the Michigan State/Notre Dame/JINA
Participation Group, Johns Hopkins University, Lawrence Berkeley
National Laboratory, Max Planck Institute for Astrophysics, Max Planck
Institute for Extraterrestrial Physics, New Mexico State University,
New York University, Ohio State University, Pennsylvania State
University, University of Portsmouth, Princeton University, the
Spanish Participation Group, University of Tokyo, University of Utah,
Vanderbilt University, University of Virginia, University of
Washington, and Yale University.
\bibliographystyle{mn2e}
\bibliography{refSC}

\label{lastpage}

\end{document}